\documentclass[12pt,preprint]{aastex}
\usepackage{amsmath}
\def\f{\frac}

\slugcomment{Accepted for publication in ApJL}
\shorttitle{Late-time afterglows of GRBs}
\shortauthors{Urata et al.}
\begin{document}
\title{Testing the External Shock Model of Gamma-Ray Bursts
using the Late-Time Simultaneous Optical and X-ray Afterglows}
\author{
Yuji~\textsc{Urata}\altaffilmark{1}, 
Ryo~\textsc{Yamazaki}\altaffilmark{2}, 
Takanori~\textsc{Sakamoto}\altaffilmark{3},
Kuiyun~\textsc{Huang}\altaffilmark{4},
Weikang~\textsc{Zheng}\altaffilmark{5},
Goro~\textsc{Sato}\altaffilmark{3},
Tsutomu~\textsc{Aoki}\altaffilmark{6}, 
Jinsong~\textsc{Deng}\altaffilmark{5},
Kunihito~\textsc{Ioka}\altaffilmark{7},
WingHuen~\textsc{Ip}\altaffilmark{8},
Koji~S.~\textsc{Kawabata}\altaffilmark{9},
YiHsi~\textsc{Lee}\altaffilmark{8},
Xin~\textsc{Liping}\altaffilmark{5},
Hiroyuki~\textsc{Mito}\altaffilmark{6},
Takashi~\textsc{Miyata}\altaffilmark{6},
Yoshikazu~\textsc{Nakada}\altaffilmark{6},
Takashi~\textsc{Ohsugi}\altaffilmark{2,9},
Yulei~\textsc{Qiu}\altaffilmark{5},
Takao~\textsc{Soyano}\altaffilmark{6},
Kenichi~\textsc{Tarusawa}\altaffilmark{6},
Makoto~\textsc{Tashiro}\altaffilmark{1},
Makoto~\textsc{Uemura}\altaffilmark{9},
Jianyan~\textsc{Wei}\altaffilmark{5},
and
Takuya~\textsc{Yamashita}\altaffilmark{8}
}

\altaffiltext{1}
{Department of Physics, Saitama University, Shimo-Okubo, Saitama,  
338-8570, Japan;
urata@heal.phy.saitama-u.ac.jp}
\altaffiltext{2}
{Department of Physics, Hiroshima University, Higashi-Hiroshima,
Hiroshima 739-8526, Japan;
ryo@theo.phys.sci.hiroshima-u.ac.jp
}
\altaffiltext{3}
{NASA Goddard Space Flight Center, Greenbelt, MD 20771, USA}
\altaffiltext{4}{Academia Sinica Institute of Astronomy and 
Astrophysics, Taipei 106, Taiwan, Republic of China}
\altaffiltext{5}{National  Astronomical Observatories, Chinese Academy of 
Sciences, Beijing 100012, China}
\altaffiltext{6}{Kiso Observatory, Institute of Astronomy, The University 
of Tokyo, Kiso-muchi, Kiso-gun, Nagano 397-0101, Japan}
\altaffiltext{7}{Departments of Physics, Kyoto University,
Kitashirakawa, Sakyo-ku, Kyoto 606-8602, Japan}
\altaffiltext{8}{Institute of Astronomy, National Central University, 
Chung-Li 32054, Taiwan, Republic of China}
\altaffiltext{9}
{Astrophysical Science Center, Hiroshima University, Higashi-Hiroshima, 
Hiroshima 739-8526, Japan}

\begin{abstract}

We study the ``normal'' decay phase of the X-ray afterglows of gamma-ray
bursts (GRBs), which follows the shallow decay phase, using the events
simultaneously observed in the R-band.
The classical external shock model --- in which neither the delayed
energy injection nor time-dependency of shock micro-physics is
considered --- shows that the decay indices of the X-ray and R-band light
curves, $\alpha_{\rm X}$ and $\alpha_{\rm O}$, obey a certain
relation, and that in particular, $\alpha_{\rm O}-\alpha_{\rm X}$
should be larger than $-1/4$ unless the ambient density increases
with the distance from the central engine.
For our selected 14 samples, we have found that 4 events violate
the limit at more than the 3$\sigma$ level, so that a fraction 
of events are outliers of the classical
external shock model at the ``normal'' decay phase.


\end{abstract}


\section{Introduction}
\label{sec:intro}

Gamma-ray bursts (GRBs) consist of two phases: prompt GRB emission and
subsequent afterglows.  How long the prompt GRB emission lasts and
when the transition from the prompt GRB to the afterglow occurs have
been long-standing problems.  These problems are tightly related to
the mechanism of the central engine of GRBs.
The {\it Swift} satellite has brought us early, dense, and detailed
data on the afterglows of GRBs in various observation bands.  Now, 
we are entering the era of multi-wavelength observations; 
especially optical and
X-ray bands, which tell us some hints for answering the problems.

Contrary to the expectation in the pre-{\it Swift} era, 
{\it Swift} X-Ray Telescope (XRT) data have
revealed complex temporal behavior of the X-ray afterglow
\citep{burrows05,tagl05,nousek06,obrien06,will07}.
Initially, it decays very steeply, whose most popular interpretation
is the tail emission of the prompt GRB
\citep{kumar00ng,zhang06,yama06}, although other possibilities have
been proposed \citep[e.g.,][]{binbin07}.  At several hundreds of
seconds after the burst trigger, the shallow decay phase begins until
$\sim10^4$ sec, whose origin is quite uncertain
\citep[e.g.,][]{toma06,ioka06,zhang07}.
After the shallow decay phase ends, the X-rays subsequently decays
with the decay index usually steeper than unity, which was expected in
the pre-{\it Swift} era. This decay behavior can be well explained 
by {\it the classical external shock model} \citep{sari98}, in which 
neither the delayed energy injection nor time-dependency of shock 
micro-physics is considered.  Hence this phase is sometimes called 
{\it the normal decay phase}.

However, as the number of the X-ray observations increases, it is
getting suspicious that the normal decay phase arises from the
external shock.
In the steep and the shallow decay phases, the X-ray light curves
sometimes possess large bumps, called {\it the X-ray flares}
\citep{chincarini07,falcon07}, and/or dips that cannot be explained by
the external shock model \citep{ioka05}.  Furthermore, for an extreme
example, GRB~070110 showed a rather complex X-ray afterglow with a
sudden drop at $\sim2\times10^4$~sec after the burst trigger as the
end of the shallow decay phase \citep{troja07}.  These observational
facts may tell us that the steep and the shallow decay phases are likely
due to late internal dissipation of the energy produced by the
long-acting central engine.
On the other hand, the X-ray spectrum remains unchanged across the
shallow-to-normal transition \citep{nousek06}, which may imply that
the shallow and the normal decay phases are of the same origin.
Therefore, it might be that the normal phase comes from the internal
energy dissipation.

The observed optical afterglow is also complicated and in an early
epoch ($\lesssim10^3$~sec), there is a diversity
\citep{zhang07,doi07}.  
On the other hand,  
it was found in the pre-{\it Swift} era that for almost all events, 
the behavior at $\gtrsim0.1$~day after the burst could be well 
explained by the {\it classical} external shock
model \citep[e.g.,][]{pana01,urata03}, although some events showed complex
light curves with dips and/or bumps \citep[e.g.,][]{holl03,lipk04,urata07b}.
This epoch corresponds to the normal decay phase of the X-ray afterglow.
Those previous studies are mainly based on the optical bands, because
the X-ray observation was sparse at that time.
In the {\it Swift} era, we 
are starting to have the simultaneous optical and X-ray afterglow 
data in the epoch $\gtrsim0.1$ day after the burst thanks to the 
rapid and the dense X-ray observation by the XRT.

In this Letter, we study  the normal decay phase of the X-ray
afterglows simultaneously observed in optical R-bands, and investigate
whether it is consistent with the classical external shock model or
not.  We perform a simple test using the optical and the X-ray decay
indices, $\alpha_{\rm O}$ and $\alpha_{\rm X}$, where we use a
notation, $F_\nu\propto t^{-\alpha}\nu^{-\beta}$.
For example, in the classical external shock model with uniform ISM,
they are related to the power-law index of the electron distribution,
$p(>2)$, as $\alpha_{\rm O}=3(p-1)/4$ and $\alpha_{\rm X}=(3p-2)/4$,
respectively, since the cooling frequency $\nu_{\rm c}$ usually lies
between the optical and X-ray bands \citep{sari98}.
Eliminating $p$, we obtain $\alpha_{\rm O}-\alpha_{\rm X}=-1/4$.
Similarly, for the wind environment, we derive $\alpha_{\rm
O}-\alpha_{\rm X}=1/4$ \citep{chev00}.
These relations between $\alpha_{\rm O}$ and $\alpha_{\rm X}$ are also
valid in the case of $1<p<2$ \citep{dai01}.
Therefore, through the relation between $\alpha_{\rm O}$ and
$\alpha_{\rm X}$, one can test the classical external shock model.
In the pre-{\it Swift} era, similar study has been done for 
{\it BeppoSAX} GRBs \citep{depasquale2006b}. However, 
compared with the {\it Swift} GRBs,
their X-ray data were not well enough to identify the normal decay phase 
and to determine the decay index with small uncertainties. 
We can now obtain more dense X-ray and optical data and can determine
$\alpha_{\rm O}$ and $\alpha_{\rm X}$ with much less ambiguity.
Finally it is noted that in this Letter, we do not consider
 the spectral indices, $\beta_{\rm O}$ and $\beta_{\rm X}$, 
because they have at present large
uncertainties; $\beta_{\rm O}$ fairly depends on the assumed dust
model, and the low X-ray flux at the epoch we are interested in
makes us difficult to constrain $\beta_{\rm X}$ with precision 
which we need to test the model.

\section{Decay indices of X-ray and R-band afterglow 
in the normal decay phase}
\label{sec:index}

We consider long GRBs that are  followed-up by 
{\it Swift} XRT from the beginning of 2005 to the end of 2006.
The {\it Swift} XRT data are systematically analyzed using our
pipeline script.  The cleaned event data of the Window Timing (WT) and
the Photon Counting (PC) mode from the {\it Swift} Science Data Center
(SDC) are used in the whole process.  Although both WT and PC mode
data are processed in the pipeline, hereafter, we are only focusing on
the process of the PC mode data.  The search of the X-ray afterglow
counterpart, a construction of the X-ray light curve, and a fitting
process of the X-ray light curve and spectra are performed
automatically using the standard XRT softwares and calibration 
database (HEASoft 6.2 and CALDB 20070531).
The source region is selected as a
circle of 47$^{\prime\prime}$ radius.  The background region is an
annulus of an outer radius of 150$^{\prime\prime}$ and an inner radius
of 70$^{\prime\prime}$ excluding the background X-ray sources detected
by {\tt ximage} in circle region of 47$^{\prime\prime}$ radius.  The
light curve is binned based on the number of photons required to meet
at least 5$\sigma$ (Sakamoto et al. 2007).
We select the samples of the X-ray afterglows which have a smooth
transition from the shallow to the normal decay phases at
$\gtrsim10^3$~sec.
Samples with X-ray flares have been excluded.
Then, we find
the start time of the normal decay phase of the X-ray afterglow
($\alpha_{\rm X} \gtrsim 1$), and extract
events in which well-sampled R-band light curves are available during
the normal decay phase.

The light curve data in the R-band are published in literatures or
observed by the East Asian GRB Follow-up Network \citep[{{\it
    EAFON}};][]{urata05}\footnote[1]{In this paper, the samples are
  mainly taken using Kiso 1.05m Schmited telescope\citep{urata03},
  Lulin One-meter telescope\citep{lot} and Xinglong 0.8m
  telescope\citep{xinglong}.} and the KANATA telescope.  The R-band
data taken by us are processed as in the following.
A standard routine, including bias subtraction, dark subtraction, and
flat-fielding corrections with appropriate calibration data is
employed to process the data using IRAF.  Flux calibrations are
performed using the APPHOT package in IRAF, referring to the standard
stars suggested by \citet{land92}.  For each data set, the
one-dimensional aperture size is set to 4 times as large as the
full-width at half maximum of the objects.  The magnitude of error for
each optical image is estimated as $\sigma_{\rm e}^{2}=\sigma_{\rm
ph}^{2} + \sigma_{\rm sys}^{2}$, where $\sigma_{\rm ph}$ represents
the photometric errors for each afterglow, estimated from the output
of IRAF PHOT, and $\sigma_{\rm sys}$ is the photometric calibration
error estimated by comparing our instrumental magnitudes.
When we combine data which are obtained at several different sites, 
we re-calibrate each data set by
our photometric manner 
\citep[e.g.,][]{huang07,urata07,urata03}. 
These efforts decrease systematic differences and
yield realistic light curves.

\begin{figure}
\epsscale{1} \plotone{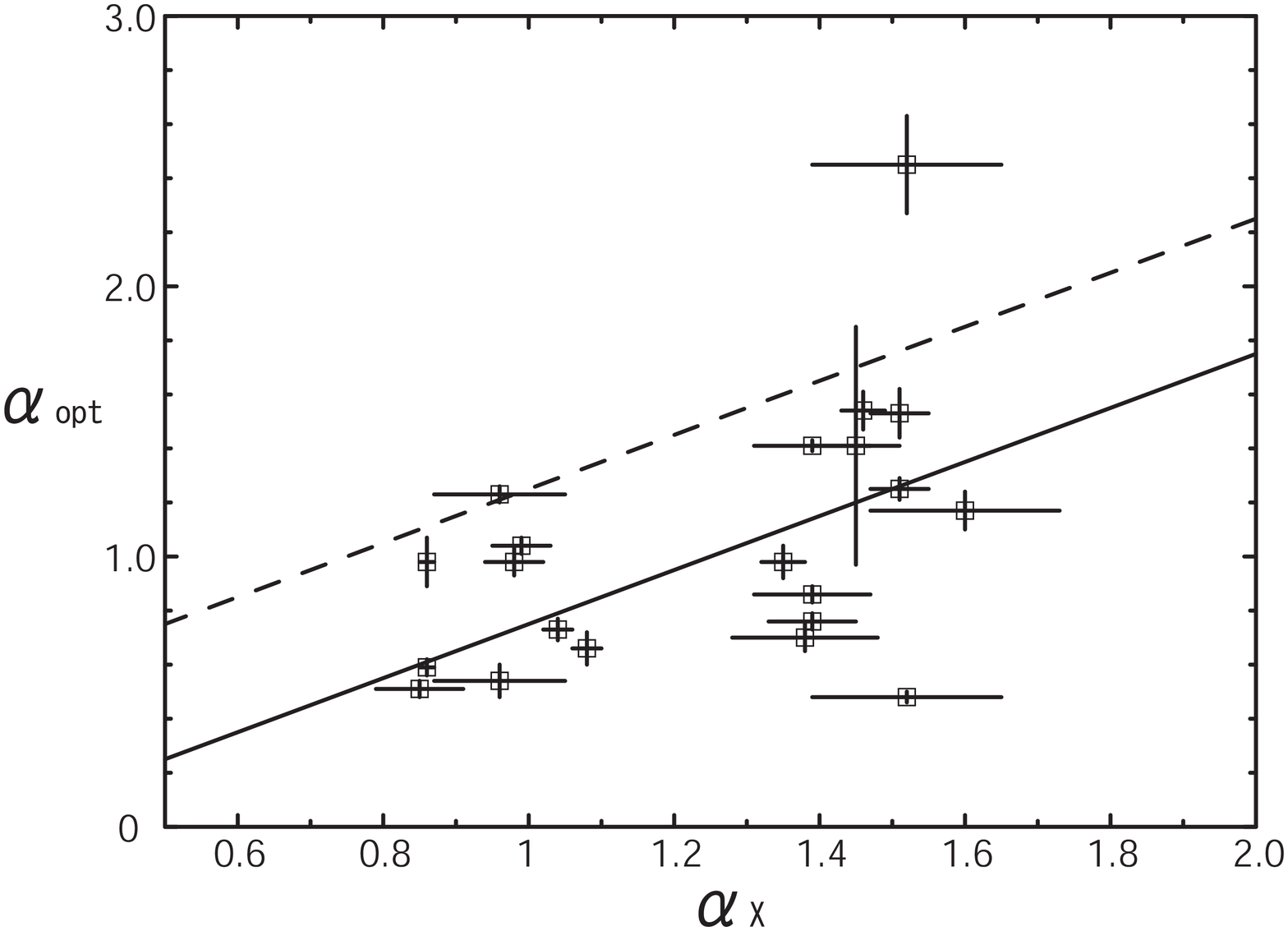}
\caption{ The R-band decay index $\alpha_{\rm O}$ as a function of the
X-ray decay index $\alpha_{\rm X}$ in the normal decay phase.  The classical
external shock model predicts $\alpha_{\rm O}-\alpha_{\rm X}=-1/4$
(solid line) and $1/4$ (dashed line) for the uniform ISM ($s=0$) and
for the wind medium ($s=2$) cases, respectively.  }
\label{fig1}
\end{figure}

There are 14 GRBs which have a good coverage with both X-ray and
optical bands at the normal decay phase.  
Among them, optical data of 11 events have been already published 
in literatures.
For unpublished data obtained by EAFON, detailed light curves in the 
X-ray and the optical bands are presented elsewhere 
(Urata et al. 2007, in preparation).
For those samples, 
we identify the normal decay phase that is well described by a single
power-law decay model and derive $\alpha_{\rm X}$.
During the phase, we
find that the optical light curves are well fitted with a single power-law model
in the time interval shown in Table~\ref{tbl2} in which
the decay index, $\alpha_{\rm O}$, is determined.
All results are summarized in Table~\ref{tbl2}.
Figure~\ref{fig1} shows $\alpha_{\rm O}$ as a function of $\alpha_{\rm X}$,
while Figure~\ref{fig2} shows the value of $\alpha_{\rm O}-\alpha_{\rm X}$ for
each event.
The quoted errors in this Letter are at the 1 $\sigma$ confidence 
level.  

\section{Results and Discussion}
\label{sec:dis}

\begin{figure}
\epsscale{1} \plotone{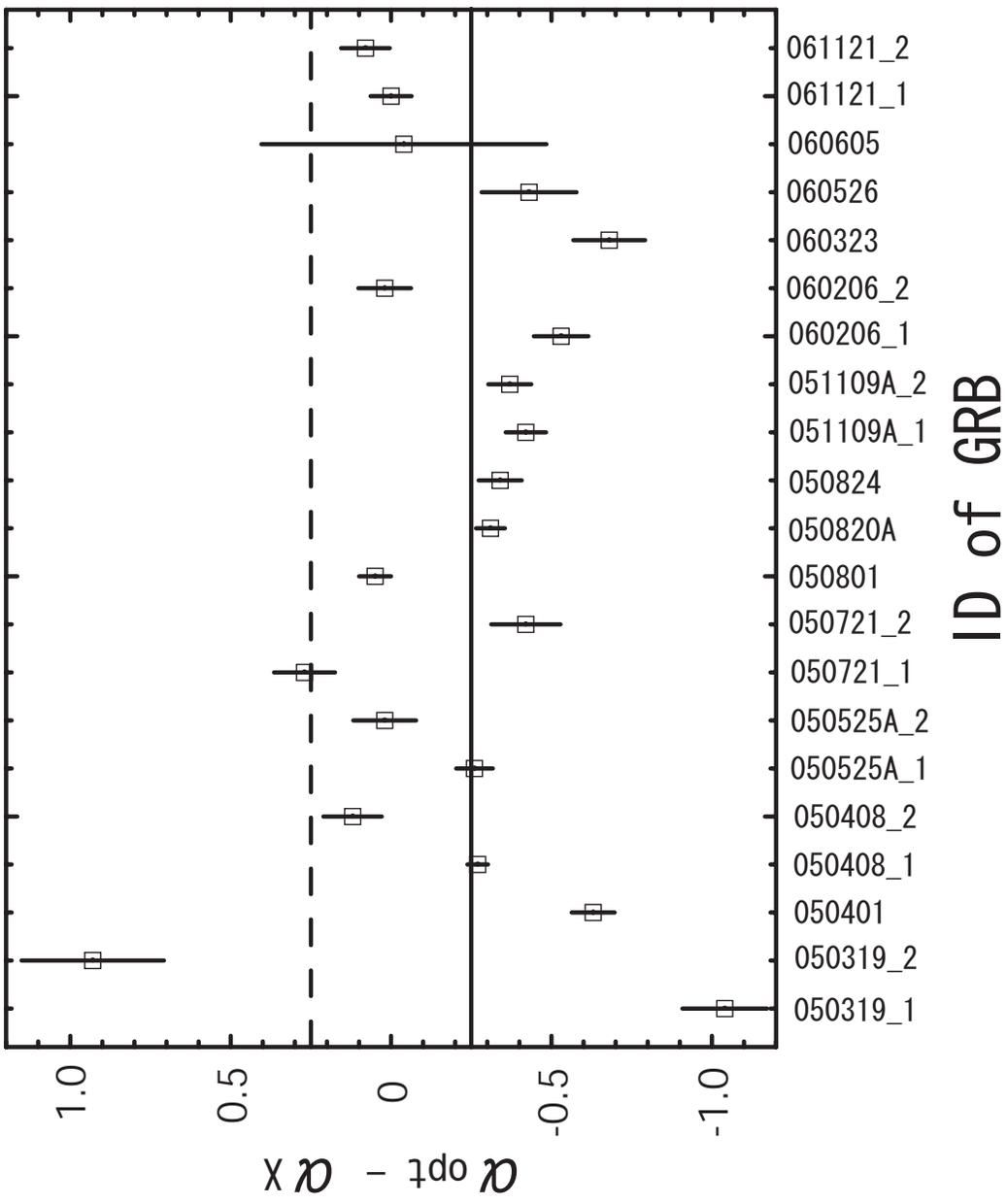}
\caption{
The range of $\alpha_{\rm O}-\alpha_{\rm X}$ for
individual events.
The meanings of the solid and the dashed lines are
the same as those in Fig.~1.
}
\label{fig2}
\end{figure}

Let us consider the case of the minimum frequency $\nu_{\rm m}$
smaller than the R-band frequency $\nu_{\rm R}$ 
($\nu_{\rm m}<\nu_{\rm R}$), which is a reasonable assumption for 
several bright bursts in the pre-{\it Swift} era.  
In this case, the spectral index 
of the optical afterglow is positive, $\beta_{\rm O}>0$,
which is consistent with the previous observational results
\citep[see Table~2 of][]{kann06}.
The decay and the spectral indices are calculated
as shown in Table~\ref{Table:index} by
the classical external shock model with ambient matter density 
dependent on the radius, $n\propto r^{-s}$ where we assume $s>0$.
Since the Lorentz factor of the relativistically expanding shell
evolves with the observer time as $\Gamma\propto t^{-\f{3-s}{8-2s}}$,
$s<3$ is needed in order for the shell to decelerate.
If $\nu_{\rm m}<\nu_{\rm R}<\nu_{\rm c}<\nu_{\rm X}$,
we derive 
\[
\alpha_{\rm O}-\alpha_{\rm X}=-\f{1}{4} + \f{s}{8-2s}~~,
\]
which is valid for $1<p<2$ or $2<p$,
so that $\alpha_{\rm O}-\alpha_{\rm X}$ ranges between
$-1/4$ and $5/4$ if $0<s<3$.
For the cases of 
$\nu_{\rm m}<\nu_{\rm R}<\nu_{\rm X}<\nu_{\rm c}$ or
$\nu_{\rm m}<\nu_{\rm c}<\nu_{\rm R}<\nu_{\rm X}$,
$\alpha_{\rm O}-\alpha_{\rm X}$ should be zero.
As can be seen in Fig.~\ref{fig2}, among 14 events considered in this
paper, 4  events (GRB~050319, 050401, 060206, 060323) are below the line 
$\alpha_{\rm O}-\alpha_{\rm X}=-1/4$
at more than the 3$\sigma$ level,
so that a fraction of
bursts are outliers of the classical external shock model at the
normal decay phase.
\citet{depasquale2006b} have performed similar study and found
two out of 12 events have $\alpha_{\rm O}-\alpha_{\rm X}$
significantly below $-1/4$, which was
 roughly consistent with our result
(see Table~5 of their paper).

\citet{liang07} studied the $\alpha_{\rm X}$--$\beta_{\rm X}$ relation
of the normal decay phase and found that there are several outliers of the 
classical external shock model.
Their outliers have large $\alpha_{\rm X}>2$.
In our sample, however, outliers of $\alpha_{\rm O}$--$\alpha_{\rm X}$
relation exists even if their $\alpha_{\rm X}$ of around 1.5,
and their $\alpha_{\rm X}$--$\beta_{\rm X}$ relations are consistent
with the classical external shock model 
\citep[see Fig.~5 of][]{liang07}.
This fact, therefore,
strengthens the importance of the multi-wavelength studies
at the normal decay phase to test the classical external shock
model.

There are several possibilities leading to 
$\alpha_{\rm O}-\alpha_{\rm X}<-1/4$.
One is to consider $s<0$ case \citep[e.g.][]{yost2003}.
If $s\lesssim-4$, then $\alpha_{\rm O}-\alpha_{\rm X}\lesssim-0.5$,
however, there is no theoretical reason to consider such a steeply
rising profile.
Another is to
consider the delayed energy injection \citep{rees1998} and/or 
time-variable shock micro-physics parameters \citep{yost2003}.
Here, we consider $\nu_{\rm m}<\nu_{\rm R}<\nu_{\rm c}<\nu_{\rm X}$
for simplicity.
The generalized forms of $\alpha_{\rm O}$ and $\alpha_{\rm X}$
are then derived by \citet{pana06} based on the assumptions
$E(>\Gamma)\propto\Gamma^{-e}$,
$\varepsilon_B\propto\Gamma^{-b}$,
$\varepsilon_e\propto\Gamma^{-i}$,
and $n\propto r^{-s}$, where $s<3$.
Then, from their derived formula, we obtain
\begin{eqnarray}
\alpha_{\rm O} - \alpha_{\rm X} &=&
-\f{1}{4} + \f{s}{8-2s} \nonumber \\
&& -\f{3-s}{4(e+8-2s)}
\left( \f{4-3s}{4-s}e+3b \right) ~~,
\label{eq:relation}
\end{eqnarray}
which is independent of $i$ and $p$.
We find that $\alpha_{\rm O}-\alpha_{\rm X}<-1/4$
is achieved if $e+3b>0$ for the uniform ISM case ($s=0$),
or if $b-e>8/3$ for the wind medium case ($s=2$).
Note that these cases have been discussed for the pre-{\it Swift} GRBs
\citep{piro1998,yost2003,corsi2005}.

Another possibility to explain the outliers may be that the X-ray
flares superposing on the X-ray afterglow could steep the apparent
decay index of the X-ray. X-ray flares are usually more active in the
initial phase, so that they may enhance the early X-ray flux. In this
case the X-ray flare should not be spiky but relatively smooth, 
and the late afterglow is just an ordinary afterglow.

Although the external shock model is still viable,
the afterglow emissions of outliers may be capable of the
internal shock origin.  Such a possibility has been proposed by
\citet{ghise07}.  Then, the optical and X-ray emission in the late
phase are of different origins.  It is also possible in this model that a
chromatic break occurs at $\sim1$~day after the burst, which was
believed to be achromatic in the pre-{\it Swift} era 
and to be caused by the jet collimation effects
\citep{pana06,huang07,sato07}.
Or a cannonball model may account for our outliers \citep{dado2007}.

\acknowledgements
We would like to thank 
Drs.~Shiho~Kobayashi and Bing~Zhang, and the anonymous referee
 for useful comments.
This work was supported in part by Grants-in-Aid for Scientific Research
of the Japanese Ministry of Education, Culture, Sports, Science,
and Technology 18001842 (Y.~U.), 18740153 (R.~Y.).
This work was also supported by NSC 93-2752-M-008-001-PAE and NSC
93-2112-M-008-006.
T.S. was supported by an appointment of the NASA Postdoctoral Program
at the Goddard Space Flight Center, administered by Oak Ridge
Associated Universities through a contract with NASA.
This paper was inspired through the discussion during the workshop
``Implications of {\it Swift}'s Discoveries about Gamma-Ray Bursts''
at the Aspen Center for Physics.


\begin{deluxetable}{lcccccl}
\tabletypesize{\scriptsize}
\rotate
\tablecaption{ 
X-ray and optical temporal decay indices during X-ray normal decay phase.
\label{tbl2}
}
\tablewidth{0pt}
%
\tablehead{
\colhead{GRB} & 
\colhead{Normal decay phase [sec]\tablenotemark{a}} &
\colhead{Optical period [sec]\tablenotemark{b}} &
\colhead{$\alpha_{\rm X}$} &
\colhead{$\alpha_{\rm O}$}  &
\colhead{$\alpha_{\rm O}-\alpha_{\rm X}$} &
\colhead{References\tablenotemark{c}}  
}
\startdata
050319\_1  & $4.8\times 10^4$ -- $2.0\times 10^6$ & $1.3\times10^5$ --  $4.1\times10^5$ & 
$1.52 \pm 0.13$ & $ 0.48 \pm  0.02$ & $-1.04 \pm  0.13 $ & (1)\\
050319\_2  & $4.8\times 10^4$ -- $2.0\times 10^6$ & $4.1\times10^5$ --  $9.9\times10^5$ & 
$1.52 \pm 0.13$ & $ 2.45 \pm  0.18$ & $ 0.93 \pm  0.22 $ & (1)\\
050401     & $3.4\times 10^3$ -- $6.3\times 10^5$ & $3.5\times10^3$ --  $1.4\times10^5$ & 
$1.39 \pm 0.06$ & $ 0.76 \pm  0.03$ & $-0.63 \pm  0.07 $ & (2)\\
050408\_1  & $2.6\times 10^3$ -- $3.2\times 10^6$ & $3.4\times10^3$ --  $4.6\times10^4$ & 
$0.86 \pm 0.01$ & $ 0.59 \pm  0.03$ & $-0.27 \pm  0.03 $ & (1), (3)\\
050408\_2  & $2.6\times 10^3$ -- $3.2\times 10^6$ & $6.2\times10^4$ --  $3.0\times10^5$ & 
$0.86 \pm 0.01$ & $ 0.98 \pm  0.09$ & $ 0.12 \pm  0.09 $ & (1), (3)\\
050525A\_1 & $3.1\times 10^3$ -- $2.7\times 10^6$ & $3.1\times10^3$ --  $5.7\times10^4$ & 
$1.51 \pm 0.04$ & $ 1.25 \pm  0.04$ & $-0.26 \pm  0.06 $ & (1), (4), (5)\\
050525A\_2 & $3.1\times 10^3$ -- $2.7\times 10^6$ & $6.3\times10^4$ --  $4.6\times10^5$ & 
$1.51 \pm 0.04$ & $ 1.53 \pm  0.09$ & $ 0.02 \pm  0.10 $ & (1), (4), (5)\\
050721\_1  & $2.3\times 10^3$ -- $3.4\times 10^6$ & $2.3\times10^3$ --  $7.9\times10^3$ & 
$0.96 \pm 0.09$ & $ 1.23 \pm  0.03$ & $ 0.27 \pm  0.09 $ & (6)\\
050721\_2  & $2.3\times 10^3$ -- $3.4\times 10^6$ & $7.9\times10^3$ --  $2.5\times10^5$ & 
$0.96 \pm 0.09$ & $ 0.54 \pm  0.06$ & $-0.42 \pm  0.11 $ & (6)\\
050801     & $6.5\times 10^2$ -- $3.0\times 10^5$ & $7.2\times10^2$ --  $9.5\times10^3$ & 
$0.99 \pm 0.04$ & $ 1.04 \pm  0.03$ & $ 0.05 \pm  0.05 $ & (7)\\
050820A    & $2.8\times 10^3$ -- $4.0\times 10^4$ & $3.4\times10^3$ --  $2.0\times10^4$ & 
$1.04 \pm 0.02$ & $ 0.73 \pm  0.04$ & $-0.31 \pm  0.04 $ & (8)\\
050824     & $5.9\times 10^4$ -- $2.0\times 10^6$ & $8.0\times10^4$ --  $4.5\times10^5$ & 
$0.85 \pm 0.06$ & $ 0.51 \pm  0.03$ & $-0.34 \pm  0.07 $ & (9)\\
051109A\_1 & $1.6\times 10^3$ -- $5.2\times 10^4$ & $1.6\times10^3$ --  $1.3\times10^4$ & 
$1.08 \pm 0.02$ & $ 0.66 \pm  0.06$ & $-0.42 \pm  0.06 $ & (10)\\
051109A\_2 & $5.2\times 10^4$ -- $1.4\times 10^6$ & $9.0\times10^4$ --  $1.0\times10^6$ & 
$1.35 \pm 0.03$ & $ 0.98 \pm  0.06$ & $-0.37 \pm  0.07 $ & (10)\\
060206\_1  & $2.3\times 10^4$ -- $5.4\times 10^5$ & $2.3\times10^4$ --  $2.5\times10^4$ & 
$1.39 \pm 0.08$ & $ 0.86 \pm  0.03$ & $-0.53 \pm  0.09 $ & (1), (11), (12)\\
060206\_2  & $2.3\times 10^4$ -- $5.4\times 10^5$ & $2.5\times10^4$ --  $2.0\times10^5$ & 
$1.39 \pm 0.08$ & $ 1.41 \pm  0.02$ & $ 0.02 \pm  0.08 $ & (1), (11), (12)\\
060323     & $1.1\times 10^3$ -- $2.1\times 10^5$ & $1.2\times10^3$ --  $3.0\times10^3$ & 
$1.38 \pm 0.10$ & $ 0.70 \pm  0.05$ & $-0.68 \pm  0.11 $ & (1)\\
060526     & $1.8\times 10^4$ -- $4.2\times 10^5$ & $2.0\times10^4$ --  $3.2\times10^4$ & 
$1.60 \pm 0.13$ & $ 1.17 \pm  0.07$ & $-0.43 \pm  0.15 $ & (13)\\
060605     & $5.2\times 10^3$ -- $2.7\times 10^4$ & $2.0\times10^4$ --  $2.3\times10^4$ & 
$1.45 \pm 0.06$ & $ 1.41 \pm  0.44$ & $-0.04 \pm  0.44 $ & (1)\\
061121\_1  & $2.1\times 10^3$ -- $1.7\times 10^4$ & $4.7\times10^3$ --  $1.5\times10^4$ & 
$0.98 \pm 0.04$ & $ 0.98 \pm  0.05$ & $ 0.00 \pm  0.06 $ & (14)\\
061121\_2  & $1.7\times 10^4$ -- $3.5\times 10^5$ & $7.2\times10^4$ --  $3.3\times10^5$ & 
$1.46 \pm 0.03$ & $ 1.54 \pm  0.07$ & $ 0.08 \pm  0.08 $ & (14), (15)\\
\enddata
\tablenotetext{a}
{The nomal decay phase is identified in the {\it Swift} XRT data. 
The value of $\alpha_{\rm X}$ is determined in this period.
Time zero is taken as the burst trigger time.}
\tablenotetext{b}
{The period when the optical data was taken during the normal decay phase.
The value of $\alpha_{\rm O}$ is determined in this epoch.} 
\tablenotetext{c}
{References for optical data.
(1) {\it EAFON} (for specific individual events, e.g. Huang et al. (2007), Deng et al. (2007));
(2) De Pasquale et al. (2006b);
(3) de~Ugarte Postigo et al. (2007);
(4) Klotz et al. (2005);
(5) Della~Valle et al. (2006);
(6) Antonelli et al.~(2006);
(7) Rykoff et al.~(2006);
(8) Cenko et al.~(2006);
(9) Sollerman et al.~(2007);
(10) Yost et al.~(2007);
(11) Wo{\'z}niak et al.~(2006); 
(12) Stanek et al.~(2007);
(13) Dai et al.~(2007);
(14) Uemura et al.~(2007); 
(15) Halpern et al. (2007)
}

\end{deluxetable}

\begin{deluxetable}{ccccc}
\tablecaption{ 
Spectral and decay indices ($F_\nu\propto t^{-\alpha}\nu^{-\beta}$)
predicted by the classical external shock model.
\label{Table:index}
}
\tablewidth{0pt}
%
\tablehead{
\colhead{} & 
\multicolumn{2}{c}{$1<p<2$} &
\multicolumn{2}{c}{$2<p$} \\
\colhead{} &
\colhead{$\alpha$} &
\colhead{$\beta$}  &
\colhead{$\alpha$} &
\colhead{$\beta$}  
}
\startdata
$\nu<\nu_{\rm m}$                  & 
$\f{(4s-3)p-2(s+3)}{3(p-1)(8-2s)}$ & $-\f{1}{3}$  & 
$\f{s-2}{4-s}$                     & $-\f{1}{3}$  \\
$\nu_{\rm m}<\nu<\nu_{\rm c}$      & 
$\f{(3-s)p+(6+s)}{2(8-2s)}$        & $\f{p-1}{2}$ & 
$\f{3p-1}{4}+\f{s}{8-2s}$          & $\f{p-1}{2}$ \\
$\nu_{\rm c}<\nu$                  & 
$\f{(3-s)p+2(5-s)}{2(8-2s)}$       & $\f{p}{2}$   & 
$\f{3p-2}{4}$                      & $\f{p}{2}$   \\
\enddata
\tablenotetext{1}{Note: For the case of the spherical expansion,
slow cooling, and the ambient density profile given by 
$n\propto r^{-s}$, where $0<s<3$.
The break frequency $\nu_{\rm m}$ evolves with time as 
$\nu_{\rm m}\propto t^{-3/2}$ for $2<p$  while 
 $\nu_{\rm m}\propto t^{-\f{(3-s)p+6-s}{(p-1)(8-2s)}}$
 for $1<p<2$, and $\nu_{\rm c}$ scales as
$\nu_{\rm c}\propto t^{\f{3s-4}{8-2s}}$ regardless of $p$.
}
\end{deluxetable}

\end{document}